\documentstyle[epsfig,epsf,psfig,aps,preprint]{revtex}
\tighten
\begin{document}
\draft
\title{Neutron star properties in a chiral SU(3) model}
\author{M. Hanauske, D. Zschiesche, S. Pal, S. Schramm, 
H. St\"ocker, W. Greiner}
\address{Institut f\"ur Theoretische Physik, J.W. Goethe-Universit\"at,
D-60054 Frankfurt am Main, Germany}

\maketitle

\begin{abstract}
We investigate various properties of neutron star matter within
an effective chiral $SU(3)_L \times SU(3)_R$ model. The
predictions of this model are compared with a Walecka-type model.
It is demonstrated that the importance of hyperon degrees are 
strongly depending on the interaction used, even if the equation 
of state near saturation density is nearly the same in both models.
While the Walecka-type model predicts a strange star core with strangeness
fraction $f_S \approx 4/3$, the chiral model allows only for
$f_S \approx 1/3$ and predicts that $\Sigma^0$, $\Sigma^+$ and 
$\Xi^0$ will not exist in star, in contrast to the Walecka-type model. 

\end{abstract}

\pacs{PACS: 26.60+c, 21.65+f, 24.10Jv}

\newpage

The internal constitution and properties of neutron stars chiefly
depend on the nature of strong interactions. The accepted underlying
theory of strong interactions, QCD, is however not solvable in the
nonperturbative regime. So far numerical solutions of QCD on a finite 
space-time lattice are unable to describe finite nuclei or infinite 
nuclear matter \cite{walebuch}. As an alternative approach several
effective models of hadronic interactions have been proposed 
\cite{sero86,bogu,fpw}. Especially the Walecka model (QHD) and its 
nonlinear extensions have been quite successful and widely used for the 
description of hadronic matter and finite nuclei. These models are 
relativistic quantum field theories of baryons and mesons, but they do 
not consider essential features of QCD, like approximate 
$SU(3)_R\times SU(3)_L$ chiral symmetry or broken scale invariance. The 
Nambu--Jona-Lasinio (NJL) model \cite{njl61a,reber96a} is an 
effective theory which has these features of QCD implemented but it lacks 
confinement and thereby fails to describe finite nuclei and nuclear matter.
The chiral SU(3) models, for example the linear SU(3)-$\sigma$ model
\cite{Torn} have been quite successful in modeling meson-meson interactions.
The $K-N$ scattering data has been well reproduced using the chiral effective
SU(3) Lagrangian \cite{Waas,Ramo}. However, all these chiral models lack the
feature of including the nucleon-nucleon interaction on the same
chiral SU(3) basis and therefore do not provide a consistent 
extrapolation to moderate and high densities relevant to the interior of
a neutron star.

This has lead us to construct a QCD-motivated chiral $SU(3)_L \times SU(3)_R$ 
model as an effective theory of strong interactions, which implements the 
main features of QCD. The model has been found to describe reasonably well,
the hadronic masses of the various SU(3) multiplets, finite nuclei,
hypernuclei and excited nuclear matter \cite{paper2,paper3}. 
The basic assumptions in the present chiral model are: (i) The 
Lagrangian is constructed with respect to the nonlinear realization of
 chiral $SU(3)_L \times SU(3)_R$
symmetry; (ii) The masses of the heavy baryons and mesons are 
generated by spontaneous symmetry breaking; (iii) The masses of the 
pseudoscalar mesons are generated by
explicit symmetry breaking, since they are the Goldstone bosons of the 
model; (iv) A QCD-motivated field $\chi$ enters, which describes the 
gluon condensate (dilaton field) \cite{Schec}; (v) Baryons and mesons 
are grouped according to their quark structure.

In this letter we investigate the composition and structure of
neutron star matter with hyperons in this chirally invariant model.
The total Lagrangian of the chiral $SU(3)_L \times SU(3)_R$ model for 
neutron star matter can be written in the mean-field approximation as
(for details see Ref. \cite{paper3})
\begin{equation}
{\cal L} = {\cal L}_{\rm kin} + {\cal L}_{\rm BM} + {\cal L}_{\rm BV} + 
{\cal L}_{\rm vec} + {\cal L}_0 + {\cal L}_{\rm SB} + {\cal L}_{\rm lep} ~,
\end{equation}
where
\begin{eqnarray}
{\cal L}_{\rm BM}+{\cal L}_{\rm BV} &=& -\sum_{i} \overline{\psi}_i \left[
m^*_i + g_{i \omega}\gamma_0 \omega^0 + g_{i \phi}\gamma_0 \phi^0 
+ g_{N\rho} \gamma_0 \tau_3 \rho_0 \right] \psi_{i} ~, \nonumber \\
{\cal L}_{\rm vec} &=& \frac{1}{2} m_{\omega}^2\frac{\chi^2}{\chi_0^2}
\omega^2 + \frac{1}{2}  m_\phi^2\frac{\chi^2}{\chi_0^2} \phi^2 
+ \frac{1}{2} \frac{\chi^2}{\chi_0^2} m_{\rho}^{2}\rho^2
+ g_4^4 (\omega^4 + 2 \phi^4 + 6 \omega^2 \rho^2 +\rho^4 ) ~, \nonumber \\
{\cal L}_0 &=& -\frac{1}{2} k_0 \chi^2 (\sigma^2+\zeta^2) 
+ k_1 (\sigma^2+\zeta^2)^2 + k_2 ( \frac{ \sigma^4}{2} + \zeta^4) 
+ k_3 \chi \sigma^2 \zeta \nonumber \\
& & - k_4 \chi^4 - \frac{1}{4}\chi^4 \ln \frac{ \chi^4 }{ \chi_0^4}
+\frac{\delta}{3}\ln \frac{\sigma^2\zeta}{\sigma_0^2 \zeta_0} ~,\nonumber \\
{\cal L}_{\rm SB} &=& -\left(\frac{\chi}{\chi_0}\right)^2
\left[m_\pi^2 f_\pi \sigma + (\sqrt{2}m_K^2 f_K - \frac{1}{\sqrt{2}} 
m_{\pi}^2 f_{\pi})\zeta \right] ,\nonumber \\
{\cal L}_{\rm lep} &=& \sum_{l=e, \mu} \overline{\psi}_l
[i \gamma_\mu\partial^\mu - m_l ]\psi_l ~.
\end{eqnarray}
Here ${\cal L}_{\rm kin}$ is the kinetic energy term of the baryons and
the scalar ($\sigma, \zeta$) and vector ($\omega, \phi, \rho$) mesons.
The interaction Lagrangian of the different baryons with the various spin-0 and 
spin-1 mesons are ${\cal L}_{\rm BM}$ and ${\cal L}_{\rm BV}$, respectively.
The sum over $i$ extends over all the charge states of the baryon octet
($p,n,\Lambda,\Sigma^-,\Sigma^0,\Sigma^+,\Xi^-,\Xi^0$).
${\cal L}_{\rm vec}$ generates the masses of the spin-1 mesons through
the interactions with spin-0 mesons, and ${\cal L}_0$ gives the 
meson-meson interaction term which induce the spontaneous breaking of chiral 
symmetry. A salient feature of the model, the dilaton field $\chi$, which
can be identified with the gluon condensate, is included. It accounts for the
broken scale invariance of QCD at tree level through the logarithmic
potential. ${\cal L}_{\rm SB}$ introduces an explicit symmetry breaking
of the U(1)$_A$, the SU(3)$_V$, and the chiral symmetry. The last term
${\cal L}_{\rm lep}$ represents the free lepton Lagrangian.
The effective masses of the baryons in the nonlinear realization of chiral
symmetry are given by \cite{paper3}
\begin{eqnarray}
 m_N^* &=& m_0 -\frac{1}{3}g_{O8}^S(4\alpha_{OS}-1)
(\sqrt{2}\zeta-\sigma) \nonumber \\
m_{\Lambda}^* &=& m_0-\frac{2}{3}g_{O8}^S(\alpha_{OS}-1)
(\sqrt{2}\zeta-\sigma) \nonumber \\
m_{\Sigma}^* &=& m_0+\frac{2}{3}g_{O8}^S(\alpha_{OS}-1)
(\sqrt{2}\zeta-\sigma) \nonumber \\
m_{\Xi}^* &=& m_0+\frac{1}{3}g_{O8}^S(2\alpha_{OS}+1)
(\sqrt{2} \zeta-\sigma) ~,
\end{eqnarray} 
with $m_0=g_{O1}^S(\sqrt{2} \sigma+\zeta)/\sqrt{3}$, in the usual notation
\cite{paper3}. The parameters $g_{O1}^S$, $g_{O8}^S$ and $\alpha_{OS}$ are 
used to fit the vacuum baryon masses to their experimental values.
The thermodynamic potential of the grand canonical ensemble per unit volume 
at zero temperature for the neutron star matter can be written as
\begin{equation}
\Omega/V = -{\cal L}_{\rm vec} - {\cal L}_0 - {\cal L}_{\rm SB}
-{\cal V}_{\rm vac} - \sum_i \frac{\gamma_i}{(2\pi)^3} \int
d^3k \left[E_i^*(k) - \mu_i^*\right] 
- \frac{1}{3} \sum_l \frac{1}{\pi^2} \int
\frac{ dk \: k^4}{\sqrt{k^2 + m_l^2}} .
\end{equation}
In Eq. (4) the vacuum energy ${\cal V}_{\rm vac}$ has been subtracted.
For a given baryon species $i$, the single particle energy and chemical 
potential are respectively, 
\begin{eqnarray}
E_i^*(k) &=& \sqrt{k_i^2 + m_i^{* 2}} ~, \nonumber \\
\mu_i &=& b_i\mu_n - q_i\mu_e = \mu_i^* + g_{i\omega}\omega_0 
+ g_{i\phi}\phi_0  + g_{i\rho}I_{3i}\rho_0 ~,
\end{eqnarray}
with $\mu_i^* \equiv E_i^*(k=k_{F_i})$; $b_i$ and $q_i$ are the baryon 
number and charge of the $i$th species.  The energy density and pressure 
follows from the Gibbs-Duhem relation, 
$\varepsilon = \Omega/V + \sum_{k=i, l}\mu_k\rho_k$ and $P=-\Omega/V$.
At a given baryon density $\rho_B$, the field equations as obtained by
extremizing $\Omega/V$ are solved self-consistently in conjunction with
the charge neutrality and $\beta$-equilibrium conditions.

The parameters of the chirally invariant potential, $k_0$ and $k_2$, are
used to ensure an extremum in the vacuum, while $k_3$ is constrained
by the $\eta'$ mass, and $k_1$ is varied to give $m_\sigma = 500$ MeV.
The vacuum expectation value of the fields $\sigma$ and $\zeta$
are constrained by the pion and kaon decay constants, $\sigma_0 = f_\pi$
and $\zeta_0 = - (2f_K-f_\pi)/\sqrt{2}$. The parameters $g_{N\omega}$ and
$\chi_0$ are used to fit the binding energy of nuclear matter 
$B/A = \varepsilon/\rho_B - m_N = -16$ MeV at the saturation density 
$\rho_0 = 0.15$ fm$^{-3}$. In the present calculation we have employed the 
parameter set C1 of Ref. \cite{paper3}. The predicted values of effective nucleon
mass, incompressibility, and symmetry energy at the saturation density
are $m^*_N/m_N = 0.61$, $K=276$ MeV, and $a_{\rm sym} = 40.4$ MeV.
The remaining couplings to the strange baryons
are then determined by the additive quark model constraints:
\begin{equation}
g_{\Lambda \omega} = g_{\Sigma \omega} = 2 g_{\Xi \omega} 
= \frac{2}{3} g_{N \omega}=2 g_{O8}^V ; \qquad 
g_{\Lambda \phi} = g_{\Sigma  \phi} = \frac{g_{\Xi \phi}}{2} = 
\frac{\sqrt{2}}{3} g_{N \omega} ~.
\end{equation} 

Figure 1 shows the energy per baryon as a function of baryonic density 
$\rho_B$ for varying neutron-proton asymmetries, 
$\delta=(\rho_n - \rho_p)/\rho_B$, calculated in the chiral model.
The curve $\delta=0$ describes infinite symmetric nuclear matter with a minimum
at $\rho_0$. With increasing asymmetry, ($\delta>0$) the binding energy 
decreases and the saturation density is shifted to lower values. The
binding in nuclear matter for small values of $\delta$ stems from the isospin 
symmetric nuclear forces. At asymmetries $\delta \geq 0.84$ (i.e. a neutron
to proton ratio $>11$), the system starts
to become unbound even at the low density regimes. The stiffest equation
of state (EOS) is obtained for pure neutron matter with $\delta=1$. Due to
the $\beta$-equilibrium conditions, the EOS for neutron star matter
(labeled NS) composed of neutrons, protons and electrons ($npe$)
is softer as compared to pure neutron matter. The gravitational attraction 
provides the necessary binding of neutron stars. The present results from 
the chiral model corroborate those obtained in Walecka-like models 
\cite{glen,Weber} and in the relativistic Brueckner-Hartree-Fock 
calculations \cite{Engv}.

Let us now discuss the inclusion of hyperons. With the choice of the 
parameter set discussed above, the chiral model
is found to produce unrealistically large hyperon potential depths in
nuclear matter in comparison to the experimental values of 
$U_{\Lambda,\Xi}^N \approx -28$ MeV for the $\Lambda$ and $\Xi$ particle.
 Parameter sets that reproduce reasonable values of 
$U_{\Lambda,\Xi}^N$ are, however, found to yield unsatisfactory nuclear
properties \cite{paper3}. Fortunately, explicit symmetry breaking can be
introduced in the nonlinear realization without affecting, e.g., the
partially conserved axial-vector currents relations. This allows for
the inclusion of additional terms for the hyperon-scalar meson coupling 
\cite{paper3}:
\begin{equation}
{\cal L}_{\rm hyp} = m_3 {\rm Tr} \left( {\overline \psi}\psi + 
{\overline\psi}\left[\psi,S\right] \right) {\rm Tr} \left(X - X_0\right) ~,
\end{equation}
where $X$ represents a scalar and $S_b^a = - [\sqrt{3}(\lambda_8)_b^a
- \delta_b^a ]/3$ with $\lambda$'s are the usual Gell-Mann matrices.
In the mean field approximation this leads to the following additional mass
term
\begin{equation}
\widetilde{m}_i^* = m_i^* +   a \: m_3 \left[\sqrt{2}(\delta -\delta_0)
+ (\zeta -\zeta_0) \right] ,
\end{equation}
where $m_i^*$ is given by Eq. (3). With $a = n_s$, where $n_s$ is the number 
of strange quarks in the baryon, and, 
with the parameter $m_3$ adjusted to $U_\Lambda^N = -28$ MeV, the
other hyperon potentials obtained are $U_\Sigma^N = +3.2$ MeV, and 
$U_\Xi^N \sim +30$ MeV. Since the potential of $\Xi$ in ground state 
nuclear matter is still not satisfactory, we have used $a=1$ as an 
alternative parametrization. For $U_\Lambda^N = -28$ MeV, one obtains
now $U_\Sigma^N = +3.2$ MeV, and $U_\Xi^N = -42$ MeV. We are well aware that
our choice of the parametrization of the hyperon potentials is not unique; 
the examination of different ways of generating the experimentally favored 
values for $U_\Sigma^N$ and $U_\Xi^N$ is in progress and will be reported 
in a different context. Hereafter we have employed $a=1$ in our calculations.

Let us compare the results obtained in the chiral model with that
of a Walecka-type model.
The latter model employed here has a cubic and quartic 
self-interaction for the $\sigma$ field, $U(\sigma) = g_2\sigma^3/3 
+ g_3\sigma^4/4$, introduced to get correct nuclear matter compressibility 
\cite{bogu}, and a quartic self-interaction for the vector 
field $\omega$, ${\cal L}_v = g_4 (\omega^\mu\omega_\mu)^2$. This
modification leads to a reasonable reproduction of the Dirac-Brueckner
calculation \cite{Suga}. We have used the parameter set TM1 of Ref. \cite{Suga}
which gives a saturation density and binding energy of $\rho_0 = 0.145$
fm$^{-3}$ and $B/A = -16.3$ MeV. It is to be noted that in the TM1 set,
the values of $m^*_N/m_N = 0.63$, $K=281$ MeV, and $a_{\rm sym} = 36.9$ MeV
which primarily influence the bulk properties of neutron star matter
are nearly identical to those in the chiral model. The hyperon couplings 
$\sigma$-$Y$ are obtained from a depth of $U_Y^N = -28$ MeV, while the 
$\omega$-$Y$ couplings are obtained from SU(6) symmetry relations (6).
Two additional strange mesons, $\zeta$ and $\phi$, are introduced which
couple only to the hyperons. The $\zeta$-Y couplings are fixed by the
condition $U_\Xi^\Xi \approx 2U_\Lambda^\Lambda \approx -40$ MeV 
\cite{sch94,sch96a}.

Figure 2 shows the particle fractions versus the baryonic density in
$\beta$-equilibrated matter in the chiral model (top panel) and the 
Walecka-type model, TM1 (bottom panel). With increasing density, it is 
energetically favorable for nucleons at the top of the Fermi sea to convert 
into other baryons. Note that the sequence of appearance of the hyperon
species is the same in both models. The first strange
particle to appear is the $\Sigma^-$, since its somewhat higher mass
(as compared to the $\Lambda$) is compensated by the electro-chemical potential
in the chemical equilibrium condition (5). Because of its negative
charge, charge neutrality can be achieved more economically
which causes a drop in the lepton fraction. More massive and positively
charged particles than these appear at higher densities. These are in fact
generic features found in neutron star calculations with hyperons
\cite{glen,Weber,Pra97,Pal}. Because of the
equilibrium condition (5), the threshold densities of the different hyperon
species are, however, strongly dependent on the magnitude of the scalar
and vector fields at a given density and their interactions with the baryons.
The nucleon-nucleon interaction which determines the variation of neutron
chemical potential with density also determines the threshold. The relatively
smaller attractive scalar fields and thereby larger effective mass of all
the baryons cause a significant shift of the density at which the hyperons 
($\Sigma^0, \Sigma^+, \Xi^0$) appear in the chiral model calculation. This 
enhances the saturation values of both the neutron and proton fractions 
(at a level of $\sim 30\%$) in the chiral model.
Consequently, the cores of massive stars in the TM1 model (and also 
in other Walecka-type models \cite{sch96a}) are in general predicted to be
very ($>50\%$) hyperon-rich. The chiral model, on the other hand, predicts
neutron stars with considerable smaller hyperon fractions in the core.

The composition of the star is crucial for neutrino and antineutrino
emission which can be responsible for the rapid cooling of neutron
stars via the URCA process. It was demonstrated \cite{Lat} that for a
$npe$ system rapid cooling by nucleon direct URCA process is allowed
when the momentum conservation condition $k_{F_p} + k_{F_e} \geq k_{F_n}$,
corresponding to a proton fraction $Y_p \geq 0.11$, is satisfied. The
magnitude of $Y_p$ at a given $\rho_B$, in turn, is determined mainly by 
the symmetry energy, which is nearly identical in the two models. 
Therefore, in both models this condition is satisfied at densities 
$\rho_B \stackrel{>}{\sim} 2.2 \rho_0$ thus, rapid cooling by direct URCA
process can occur.

The most profound implication of the constitution of neutron star matter
on its bulk properties are manifested in the equation of state (EOS).
The pressure $P$ versus the energy density $\varepsilon$ is displayed in 
Fig. 3 for neutron star matter in the chiral (thick lines) and 
TM1 (thin lines) model. For $npe$ stars (solid lines), although the 
incompressibility $K$ and the effective nucleon mass $m^*_N$ at the normal
nuclear matter density are similar in the two models, the EOS 
at large densities is found to be considerably softer
for the chiral model as compared to the EOS for TM1. In the latter model, the
$m^*_N$ rapidly decreases with density, consequently the EOS passes quickly
to one that is dominated by the repulsive vector mesons ($\omega$ and $\rho$)
leading to a stiffer EOS. This has a strong bearing on the mass and radius 
of such stars discussed below. The structure of static neutron stars can be 
determined by solving the Tolman-Oppenheimer-Volkoff equations \cite{Tol}. 
We use the results of Baym, Pethick and Sutherland \cite{bay71a} to describe 
the crust consisting of leptons and nuclei at the low-density ($\rho_B<0.001$ 
fm$^{-3}$) EOS. For the mid-density regime ($0.001 < \rho_B < 0.08$ fm$^{-3}$) 
the results of Negele and Vautherin \cite{nege73a} are employed.
Above this density, the EOS for the relativistic models have been adopted.
The masses $M$ of the nonrotating neutron star sequence is shown in 
Fig. 4 as a function of central energy density $\varepsilon_c$ in the 
two models. The corresponding mass-radius relationship is presented
in Fig. 5. It is observed that the stiffer EOS for a $npe$ star in the TM1
model can support a larger maximum mass of $M_{\rm max} = 2.16 M_\odot$
with a corresponding radius of $R_{M_{\rm max}} = 11.98$ km at a central 
baryonic density of $\rho_c = 6.08\rho_0$. The corresponding values 
obtained in the chiral model are $M_{\rm max} = 1.84 M_\odot$, 
$R_{M_{\rm max}} = 11.10$ km, and $\rho_c = 6.98\rho_0$. The large
difference in the $M_{\rm max}$  values obtained in the two models, with
nearly identical $K$ and $m^*_N$ values at $\rho_0$, clearly demonstrates 
that measurements of maximum masses of neutron ($npe$) stars at high 
densities cannot be used to constrain the incompressibility and the effective
nucleon mass around the nuclear matter densities. Constraints on $K$ and 
$m^*_N$ from radius measurements of massive stars will be even more uncertain, 
since about $40\%$ of these stars' radius originates from the low density EOS. 
In fact, no precise radius measurements currently exist.

When the hyperon degrees of freedom are included, the EOS (represented by
dash-dotted lines in Fig. 3) for both the models are appreciably softer
as compared to $npe$ stars. This is caused by the opening of the hyperon
degrees of freedom which relieves some of the Fermi pressure of the nucleons.
Also the decrease of the pressure exerted by the leptons (they are
replaced by negatively charged hyperons to maintain charge neutrality)
contributes to softening of the EOS. Since the threshold density for the 
appearance of the hyperons, especially ($\Sigma^0,\Sigma^+,\Xi^0$)
(see Fig. 2), are smaller in TM1 model, these stars contain more baryon 
species. This leads to an enhanced softening as compared to that in the
chiral model. In fact, both models with hyperons predict quite similar values
of pressure at moderate and high densities; the structures observed in the 
EOS (Fig. 3) correspond to the population of different hyperon species. These
should reflect both in the masses and radii of the stars. The maximum
masses and corresponding radii of stars with hyperons in the two models
are almost identical with $M_{\rm max} = 1.52 M_\odot$, 
$R_{M_{\rm max}} = 11.64$ km, and $\rho_c = 5.92\rho_0$ in the chiral model, 
while in the TM1 model, $M_{\rm max} = 1.55 M_\odot$, 
$R_{M_{\rm max}} = 12.14$ km, and $\rho_c = 5.97\rho_0$. The magnitude
of the central densities indicates that the hyperons 
($\Sigma^0,\Sigma^+,\Xi^0$) are entirely precluded in stars for the chiral 
model whereas all hyperon species appear with comparable abundances
in the TM1 model for the maximum-mass star. The strangeness fraction, 
$f_S = \sum_i |S_i|\rho_i /\rho_B$, are vastly different at the center of 
the maximum-mass stars: 0.33 vs. 0.75 in the chiral and in the TM1 model, 
respectively. Thus, models with similar nuclear matter 
incompressibilities and effective nucleon masses, leading to similar
maximum star masses and corresponding similar radii can however have a 
widely different baryonic constitution!

For progressively smaller central densities, the masses of the stars with 
hyperons are larger in the TM1 model (see Fig. 4), although the pressure
at a given density is smaller than in the
chiral model. This is because the masses of stars are determined by the
overall EOS, and the TM1 model possess a distinctively stiffer EOS 
at moderate densities (see Fig. 3). The radii of these small mass 
hyperon-rich stars are quite distinct in these two models (see Fig. 5),
because the radius
of the stars depend most sensitively on the low density behavior of the EOS.
In both models, stars of mass $1.44 M_\odot$ (corresponding to the 
lower limit imposed by the larger mass of the binary pulsar PSR 1913+16
\cite{Wei}) also contain many hyperon species with sizeable concentration.

In conclusion, we have investigated the composition and structure
of neutron star matter in a novel chiral SU(3) model, and compared its
predictions with that of a Walecka-type model based on different 
underlying assumptions. The two models with nearly identical values of 
effective nucleon mass, incompressibility, and symmetry energy at the normal 
nuclear density yield widely different maximum neutron star masses and radii. 
When the hyperon degrees of freedom are included, the maximum masses and
the corresponding radii in the two models are found to be rather similar.
However, softness of the nucleonic contribution in the chiral model 
precludes the hyperons $\Sigma^0,\Sigma^+,\Xi^0$ leading to much smaller 
hyperon abundances in these stars.

\begin{acknowledgements}
The authors are thankful to N. Glendenning, F. Weber and J. Schaffner-Bielich 
for helpful discussions. This work was funded in part by the Gesellschaft f\"ur 
Schwerionenforschung (GSI), the DFG and the Hessische Landesgraduiertenf\"orderung. 
S.P. acknowledges support from the Alexander von Humboldt Foundation.
\end{acknowledgements}




%
\begin{figure}[h]
\centerline{\mbox{
\epsfxsize=12cm\epsffile{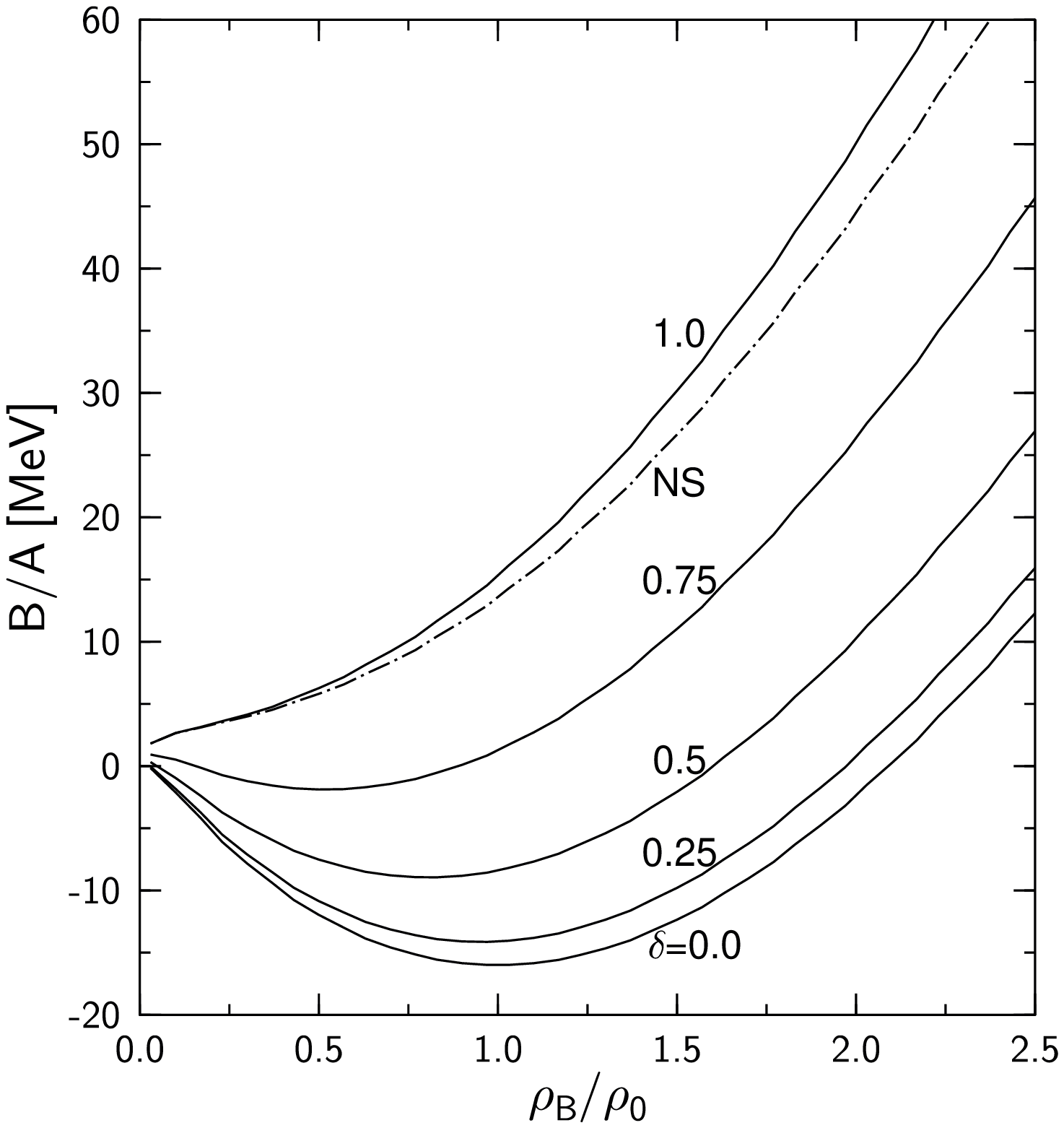}}}
\caption{The binding energy per nucleon $B/A$ versus the baryonic density
$\rho_B/\rho_0$ for different values of neutron-proton asymmetries 
$\delta=(\rho_n - \rho_p)/\rho_B$ in the chiral model. The normal nuclear 
matter density is $\rho_0=0.15$ fm$^{-3}$. The curve labeled NS describes 
a neutron star matter consisting of nucleons and electrons.}
\end{figure}
\begin{figure}[h]
\centerline{\mbox{
\epsfxsize=12cm\epsffile{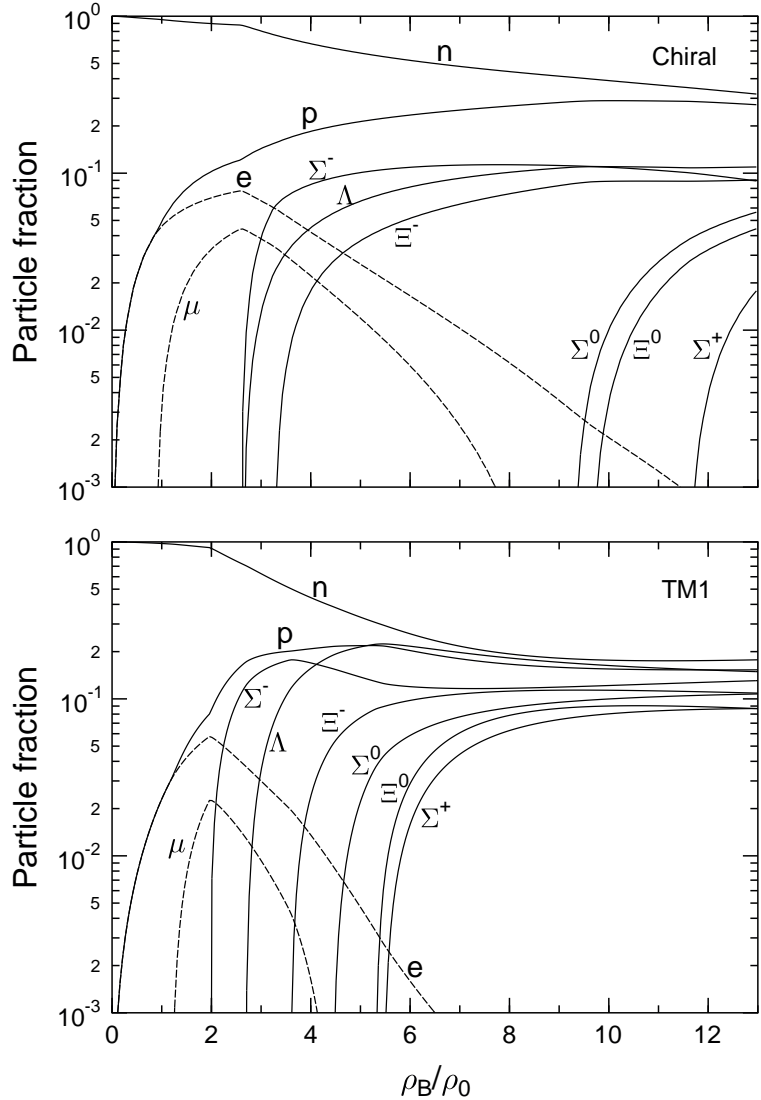}}}
\caption{The composition of neutron star matter with hyperons 
in the chiral (top panel) and in the TM1 (bottom panel) model.
The normal nuclear matter density is $\rho_0=0.15$ and 0.145 fm$^{-3}$
in the chiral and TM1 model.}
\end{figure}
\begin{figure}[h]
\centerline{\mbox{
\epsfxsize=12cm\epsffile{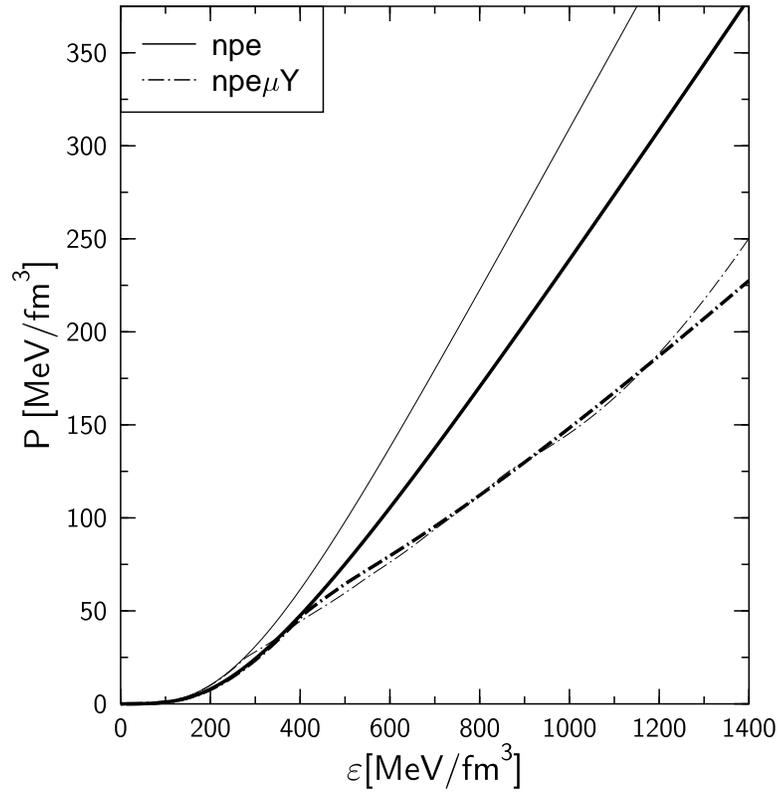}}}
\caption{The equation of state, i.e. the pressure $P$, versus the energy density
$\varepsilon$. The results are for $npe$ stars (solid lines), and for
stars with further inclusion of hyperons and muons (dash-dotted lines).
The calculations are in the chiral model (thick lines) and in the TM1
model (thin lines).}
\end{figure}
\vspace{1cm}
\begin{figure}[h]
\centerline{\mbox{
\epsfxsize=12cm\epsffile{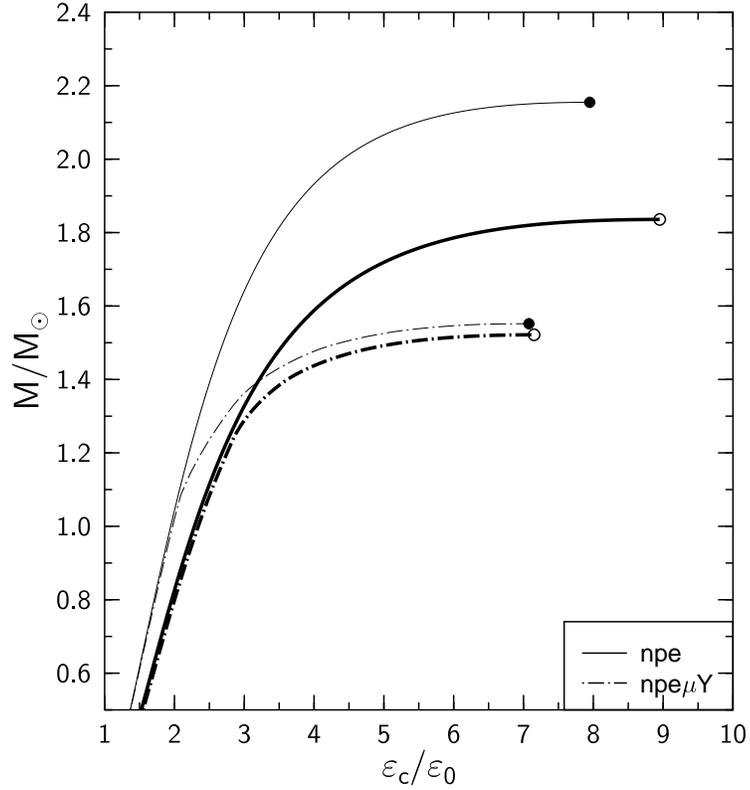}}}
\caption{The mass as a function of central energy density 
$\varepsilon_c/\varepsilon_0$ for $npe$ stars and for stars with further 
inclusion of hyperons and muons in the chiral and TM1 models.  The chiral
model results are represented by thick lines, and the TM1 results are given
by the thin lines. The circles correspond to the respective 
maximum masses. The central energy density at normal nuclear matter value is 
$\varepsilon_0 = 142.9$ and 137.0 MeV/fm$^3$ in the chiral and in the TM1 model,
respectively.}
\end{figure}
\begin{figure}[h]
\centerline{\mbox{
\epsfxsize=12cm\epsffile{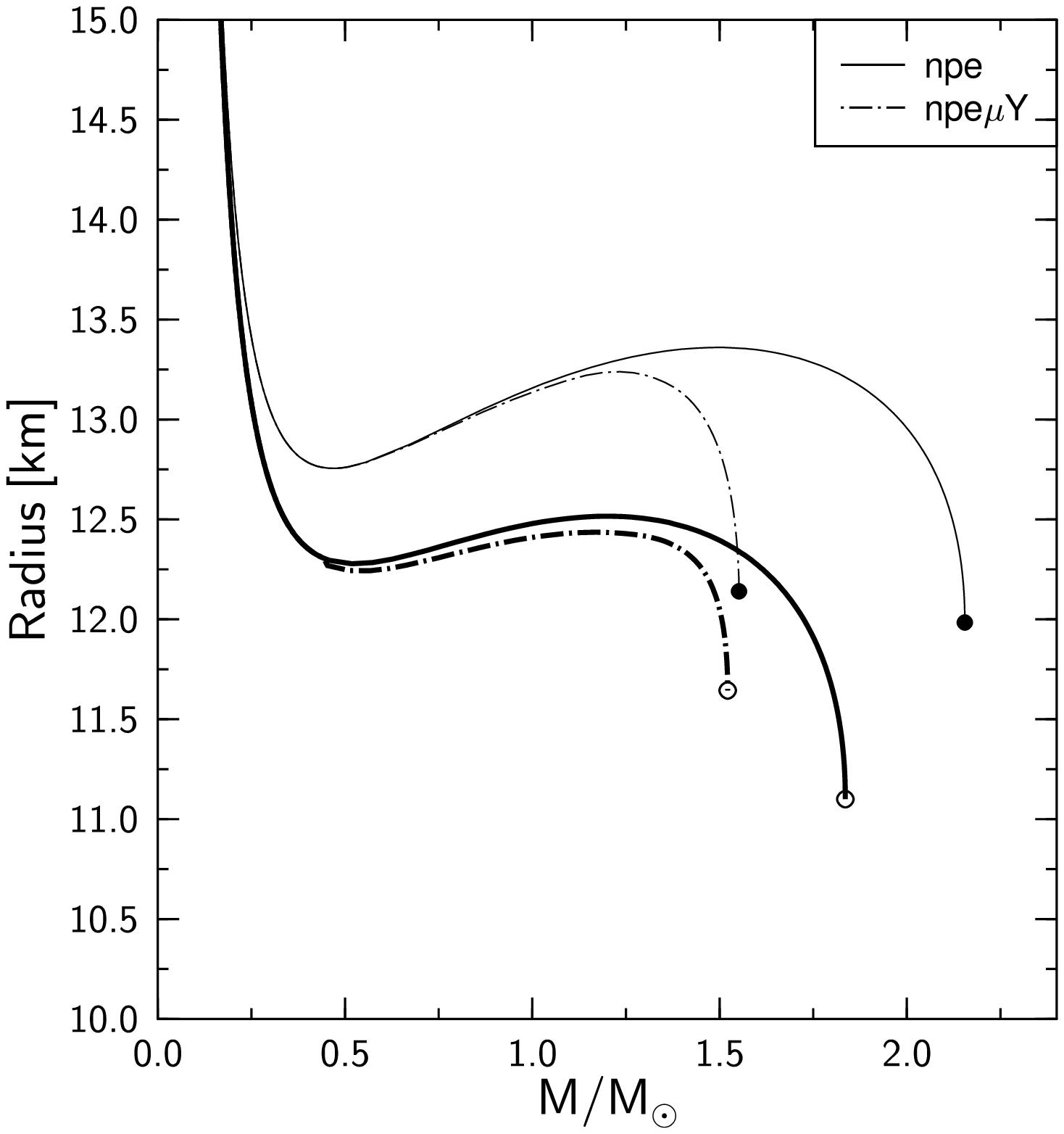}}}
\caption{The mass as a function of radius for neutron stars consisting 
of $npe$ and for stars with further inclusion of hyperons and muons in 
the chiral and TM1 model. The symbols have the same meaning as in Fig. 4.}
\end{figure}
\vspace{1cm}

\newpage
\begin{thebibliography}{99}

\bibitem{walebuch} J.D. Walecka, {\em Theoretical Nuclear and Subnuclear 
Physics} (Oxford University Press, New York, 1995).
\bibitem{sero86} B.D. Serot and J.D. Walecka, Adv. Nucl. Phys. 16 (1986) 1.
\bibitem{bogu} J. Boguta and A.R. Bodmer, Nucl. Phys. A  292 
(1977) 413;\\
J. Boguta and H. St\"ocker, Phys. Lett.  B120 (1983) 289.
\bibitem{fpw} R.J. Furnstahl, C.E. Price, and G.E. Walker, Phys. Rev. C 
36 (1987) 2590.
\bibitem{njl61a} Y. Nambu and G. Jona-Lasinio, Phys. Rev.  122  
(1961) 345; 124 (1961) 246.
\bibitem{reber96a} P. Rehberg, S. Klevansky, and J. H\"ufner, Phys. Rev. C 
53 (1996) 410.
\bibitem{Torn} N.A. T\"ornqvist, hep-ph/9711483.
\bibitem{Waas} T. Waas and W. Weise, Nucl. Phys. A 625 (1997) 287.
\bibitem{Ramo} A. Ramos and E. Oset, Nucl. Phys. A 635 (1998) 99.
\bibitem{paper2} P. Papazoglou, S. Schramm, J. Schaffner-Bielich, 
H. St\"ocker, and W. Greiner, Phys. Rev. C 57 (1998) 2576.
\bibitem{paper3} P. Papazoglou, D. Zschiesche, S. Schramm, J. Schaffner-Bielich, 
H. St\"ocker, and W. Greiner, Phys. Rev. C 59 (1999) 411.
\bibitem{Schec} J. Schechter, Phys. Rev D 21 (1980) 3393.
\bibitem{glen}N.K. Glendenning, {\em Compact Stars} (Springer, New York, 1997).
\bibitem{Weber}
F. Weber, {\em Pulsars as Astrophysical Laboratories for Nuclear and 
 Particle Physics} (IoP, Bristol, 1999).
\bibitem{Engv} L. Engvik, M. Hjorth-Jensen, E. Osnes, G. Bao, and
E. {\O}stgaard, Phys. Rev. Lett. 73 (1994) 2650.
\bibitem{Suga} Y. Sugahara and H. Toki, Nucl. Phys. A 579 (1994) 557.
\bibitem{sch94} J. Schaffner, C.B. Dover, A. Gal, D.J. Millener, C. Greiner,
and H. St\"ocker, Ann. Phys. (N.Y.)  235 (1994) 35.
\bibitem{sch96a} J. Schaffner and I. Mishustin, Phys. Rev. C 43  (1996) 1416.
\bibitem{Pra97} M. Prakash, I. Bombaci, M. Prakash, P.J. Ellis, J.M. Lattimer, 
and R. Knorren, Phys. Rep.  280 (1997) 1. 
\bibitem{Pal} S. Pal, M. Hanauske, I. Zakout, H. St\"ocker, and W. Greiner,
Phys. Rev. C 60 (1999) 015802.
\bibitem{Lat} J.M. Lattimer, C.J. Pethick, M. Prakash, and P. Haensel, 
Phys. Rev. Lett. 66 (1991) 2701.
\bibitem{Tol} R.C. Tolman, Phys. Rev.  55 (1939) 364;\\ 
J.R. Oppenheimer and G.M. Volkoff, Phys. Rev. 55 (1939) 374.
\bibitem{bay71a} G. Baym, C. Pethick, and P. Sutherland, Astrophys. J. 
170 (1971) 299.
\bibitem{nege73a} J. Negele and D. Vautherin, Nucl. Phys. A  207
(1973) 298.
\bibitem{Wei} J.M. Weisberg and J.H. Taylor, Phys. Rev. Lett. 52 
(1984) 1348.

\end{thebibliography}
\end{document}